\documentclass[useAMS,usenatbib]{mn2e}
\usepackage{placeins}
\usepackage{epsf}
\usepackage{amssymb}
\usepackage{amsmath}
\usepackage[usenames]{color}
\usepackage{graphicx}

\newcommand {\Chandra}{{\it CHANDRA }}

\newcommand{\be}{\begin{equation}}
\newcommand{\ee}{\end{equation}}

\def\***#1{\textbf{\textsf{*** #1 ***}}}

\title[X-ray image arithmetic]{Arithmetic with X-ray images of galaxy clusters: effective equation of state for small-scale perturbations in the ICM}

\author[Churazov et al.]{E.~Churazov,$^{1,2}$ P.~Arevalo$^3$, W.~Forman$^4$, C.~Jones$^4$, A.~Schekochihin$^{5,6}$, \newauthor A.~Vikhlinin,$^{4}$, I.~Zhuravleva$^{7,8}$ \newauthor \\  
$^1$ Max-Planck-Institut f\"ur Astrophysik (MPA), Karl-Schwarzschild-Strasse 1, 85741
Garching, Germany\\
$^2$ Space Research Institute (IKI), Profsoyuznaya 84/32, Moscow 117997, 
Russia\\
$^3$Instituto de F{\'i}sica y Astronom{\'i}a, Facultad de Ciencias, Universidad 
de Valpara{\'i}so, Gran Bretana N 1111, Playa Ancha,
Valpara{\'i}so, Chile\\
$^4$ Harvard-Smithsonian Center for Astrophysics, 60 Garden St.,
Cambridge, MA 02138, USA \\
$^5$Rudolf Peierls Centre for Theoretical Physics, University of Oxford, 1 Keble
 Rd, Oxford OX1 3NP, UK\\
$^6$Merton College, University of Oxford, Merton St, Oxford OX1 4JD,
UK\\
$^7$Kavli Institute for Particle Astrophysics and Cosmology, Stanford University
, 452 Lomita Mall, Stanford, California 94305-4085, USA\\
$^8$Department of Physics, Stanford University, 382 Via Pueblo Mall,
Stanford, California 94305-4060, USA\\
}

\begin{document}

\pagerange{\pageref{firstpage}--\pageref{lastpage}}

\maketitle

\label{firstpage}
\begin{abstract}
We discuss a novel technique of manipulating X-ray images of galaxy
clusters to reveal the nature of small-scale density/temperature
perturbations in the intra cluster medium (ICM). As we show, this
technique can
be used to differentiate between sound waves and isobaric
perturbations in Chandra images of the Perseus and M87/Virgo
clusters. The comparison of the manipulated images with the radio
data and with the results of detailed spectral analysis shows that
this approach successfully classifies the types of perturbations and
helps to reveal their nature. For the central regions (5-100 kpc) of the M87 and Perseus clusters
this analysis suggests that observed images are dominated by isobaric perturbations, followed by perturbations caused by bubbles of relativistic plasma and weak shocks. Such a hierarchy is best explained in a ``slow'' AGN feedback scenario, when much of the mechanical
energy output of a central black hole is captured by the bubble enthalpy that is gradually released during buoyant rise of the bubbles. 
The ``image arithmetic'' works best for prominent structure and for datasets with excellent statistics,
visualizing the perturbations with a given effective equation of state. The same approach can be extended to faint
perturbations via cross-spectrum analysis of surface brightness
fluctuations in X-ray images in different energy bands.
\end{abstract}

\begin{keywords}
\end{keywords}

%

\sloppypar

\section{Introduction}
Apart from global density and temperature radial profiles, 
fundamental information on the physics of the ICM is contained in the
small-scale deviations of the ICM properties from this global
model. Understanding the nature of the perturbations is particularly
important to constrain conduction and viscosity in hot gaseous
atmospheres of galaxies, groups, and clusters and to probe physical
mechanisms driving these perturbations.

Density and temperature perturbations of the ICM are found in
simulations \citep[e.g.,][]{2007ApJ...659..257K,2013MNRAS.428.3274Z,2013MNRAS.432.3030R} and observations \citep[e.g.,][]{1996ApJ...465L...1M}. They reflect the
non-stationary nature of a cluster and can be caused by a variety of
processes, ranging from mergers to radiative cooling of the cluster
gas. Especially rich substructure is found in the cores of relaxed
clusters, where AGN feedback plays a key role
\citep[e.g.,][]{1993MNRAS.264L..25B,2000A&A...356..788C,2000MNRAS.318L..65F,2000ApJ...534L.135M,2002ApJ...567L.115J,2007ApJ...665.1057F}.
The key question is an objective characterization of these
perturbations. Properties of the perturbations not only reflect the
driving mechanisms behind them, but also depend on the microphysics of
the ICM, in particular, on the thermal conduction and viscosity \citep[e.g.,][]{2013ApJ...764...60R,2013A&A...559A..78G,2015ApJ...798...90Z}.

One possible way of revealing the nature of the perturbations is via
measuring the correlation between density and temperature
fluctuations, to construct an effective ``equation of state'' for
the perturbations. For example, a positive correlation between temperature
and density fluctuations suggests that the gas is adiabatically
compressed \citep[e.g.,][]{2004A&A...426..387S}, while a negative
correlation hints at isobaric perturbations, associated with entropy
fluctuations
\citep[e.g.,][]{1996ApJ...472L..17M,2003ApJ...590..225C,2007ApJ...665.1057F}.
Various flavours of such analysis have been done
\citep[e.g.,][]{2001ApJ...549..228S,2004A&A...426..387S,2005A&A...442..827F,2008ApJ...687..936K,2009ApJ...700.1161G,2016ApJ...818...14A,2016MNRAS.458.2902Z,2016A&A...585A.130H}.
Here we discuss a novel technique of manipulating X-ray images of
galaxy clusters to reveal the nature of small-scale
density/temperature perturbations in the ICM.

The structure of the paper is as follows. In \S\ref{sec:tmap} we
discuss a modification of the hardness-ratio technique. We factorize  the
hardness ratio map into large-scale and small-scale
maps, using Chandra data on the Perseus cluster for illustration. In the rest of the paper we use
(and further modify) the small-scale component of the hardness ratio. In
\S\ref{sec:em} we predict the response in different energy bands to
linear perturbations of the ICM thermodynamic properties. In
\S\ref{sec:arith} we suggest a way to manipulate X-ray images in order to
suppress a particular type of perturbations and provide illustrative examples in \S\ref{sec:per}. The results are summarized in \S\ref{sec:dis}.

\section{Projected hardness-ratio and temperature maps}
\label{sec:tmap}
Generating projected temperature maps can be straightforwardly done by
partitioning the image into regions with a sufficient number of photon counts,
extracting spectra and then approximating them with a model of
optically thin plasma. This procedure can be much simplified if the
temperature variations are not very large. We describe below two
techniques that provide a quick replacement for an often lengthy direct fitting procedures.

\begin{figure*}
    \begin{minipage}{0.87\textwidth}
    \includegraphics[trim= 0mm 0cm 0mm 0cm, width=1\textwidth,clip=t,angle=0.,scale=
  1.0]{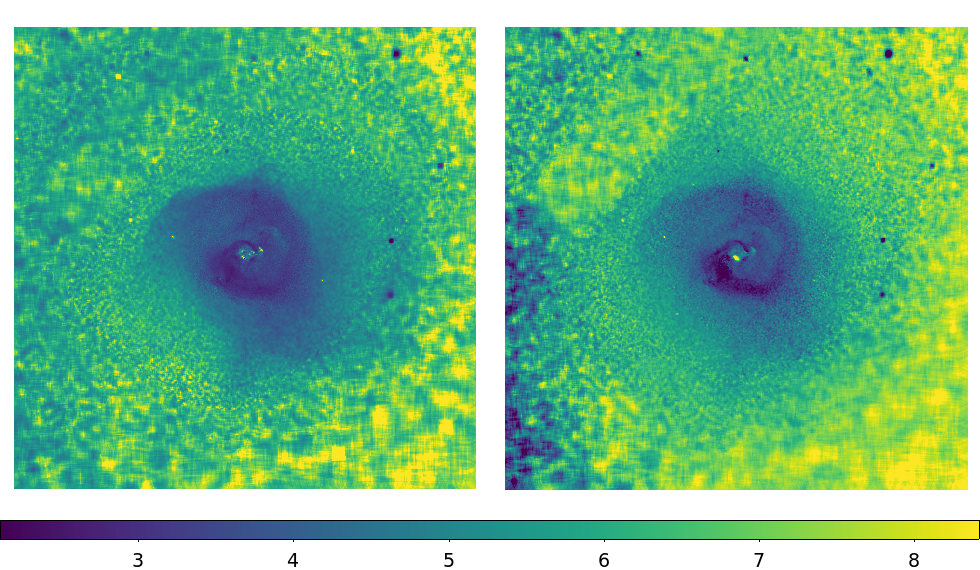}
  \end{minipage}
\caption{Projected temperature and hardness ratio maps ($16'\times16'$) of the Perseus
  cluster. {\bf Left:} The temperature map calculated from Chandra
  data using the approach described in
  \citet{1996ApJ...471..673C}. Small dark spots, e.g., in the
  upper-right corner, correspond to background AGNs that were not
  excised when constructing the temperature map. Bright yellow patches
  near the center are  due to low-energy absorption features in
  the Perseus core \citep{2000MNRAS.318L..65F}. {\bf Right:} Hardness-ratio map based on
  eq.~(\ref{eq:hr}) using the two energy bands 0.5-3.5~ keV and 3.5-7.5~keV.
\label{fig:tmap}
\label{fig:hr}
}
\end{figure*}
\begin{figure*}
    \begin{minipage}{0.87\textwidth}
    \includegraphics[trim= 0mm 0cm 0mm 0cm, width=1\textwidth,clip=t,angle=0.,scale=
  1.0]{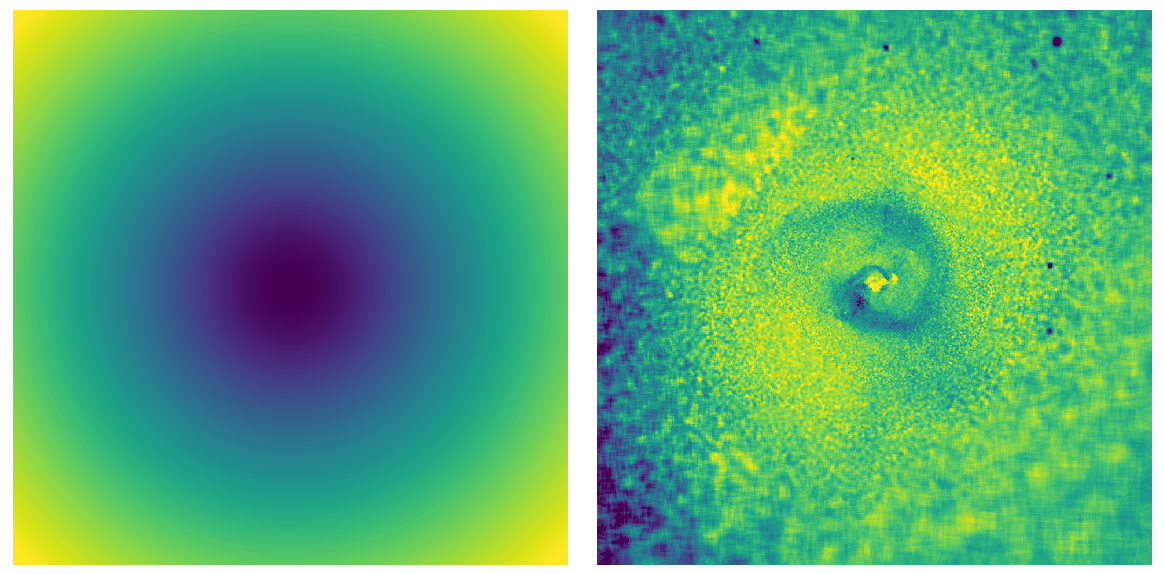}
  \end{minipage}
    \caption{Decomposition of hardness ratio $H$ into large-scale $H_g$ and small-scale $H_p$ components, $H=H_gH_p$ [see, eq.~(\ref{eq:hr}) and \S\ref{sec:mhr}]. {\bf Left:} Large scale radial dependence of the hardness ratio $H_g$,
  calculated as the ratio of best-fitting beta models for the 0.5-3.5 and 3.5-7.5 keV bands $\displaystyle H_g=\frac{I^0_h(x,y)}{I^0_s(x,y)}$. {\bf Right:} Small-scale variations of the hardness ratio $\displaystyle H_p=\left(1+\frac{\delta I_h(x,y)}{I^0_h(x,y)}- \frac{\delta I_s(x,y)}{I^0_s(x,y)}\right)$. The value of $H_p$ characterizes temperature variations, associated with departures from symmetric smooth models. The image has been adaptively smoothed with a boxcar filter.  A bright feature close to the nucleus in the right panel is caused by cold gas absorption \citep[see, e.g.,][]{2000MNRAS.318L..65F}. A boxy structure in the top-left corner corresponds to the underexposed area compared to the central region, which is covered with multiple long pointings. 
\label{fig:hrpt}
}
\end{figure*}

\subsection{Taylor expansion of the temperature dependence}
One possible way of quickly constructing a projected temperature map of a cluster \citep{1996ApJ...471..673C} is to make a Taylor
expansion of the plasma emissivity  $\epsilon(E,T)$ with respect to
small changes of temperature $T$:
\begin{eqnarray}
  \epsilon(E,T)\approx \epsilon(E,T_0)+\left.\frac{\partial \epsilon(E,T)}{\partial T}\right |_{T_0}\delta T,
  \label{eq:tlin}
\end{eqnarray}
where $E$ is the photon energy, $\delta T=T-T_0$ is a small deviation of temperature from a reference value. The functions of energy in the r.h.s., $\epsilon(E,T_0)$ and
$\displaystyle \frac{\partial \epsilon(E,T)}{\partial T}$, can be easily
generated using the Astrophysical Plasma Emission Code (APEC) \citep{2001ApJ...556L..91S} or the SPEX package \citep{1996uxsa.conf..411K}. In practice, it is convenient to replace $\displaystyle
\frac{\partial \epsilon(E,T)}{\partial T}$ with a ``macroscopic derivative''
$\displaystyle \frac{\epsilon(E,T_2)-\epsilon(E,T_1)}{T_2-T_1}$, where $T_1$ and
$T_2$ are the values of temperature bracketing the expected range of
temperature variations in a given cluster. Then eq.~(\ref{eq:tlin}) can
be re-written as
\begin{eqnarray}
  \epsilon(E,T)\approx a_1 \epsilon(E,T_1)+a_2\epsilon(E,T_2),
  \label{eq:2comp}
\end{eqnarray}
 where $\epsilon(E,T_1)$
and $\epsilon(E,T_2)$ are given and $a_1$ and $a_2$ are the two free
parameters of the model. Fitting the observed spectrum to eq.~\ref{eq:2comp} reduces to a trivial linear
problem that is computationally fast (based on the calculation of the scalar
product of observed counts with reference models $\epsilon(E,T_1)$ and
$\epsilon(E,T_2)$) and can be done in the highest resolution pixels, even if there are only a few
counts in these pixels. Once the maps of $a_1$ and $a_2$ are generated, they can be
(adaptively) smoothed and combined to determine the value of the temperature, which
in the simplest form is
\begin{eqnarray}
T=\frac{a_1T_1+a_2T_2}{a_1+a_2},
\end{eqnarray}
see \citet{1996ApJ...471..673C} for a more accurate version of this
  expression, that ensures that the value of $T$ is recovered exactly not only at the boundaries of the $[T_1:T_2]$ interval, but also at $T=(T_1+T_2)/2$. A comparison
with the results of direct fitting has shown that despite the simplicity of the method, the derived values of
temperature are accurate and have similar statistical uncertainty. The
temperature map of the Perseus cluster obtained from Chandra
observations, in this way, is shown in the left panel of Fig.~\ref{fig:tmap}.

As a caveat, we mention that this procedure assumes that all other
parameters, e.g., abundance of heavy elements or low energy photoelectric absorption,
do not affect the hardness rato. In fact, in the very core of the Perseus cluster there are clear signes of strong low-energy absorption features \citep[see, e.g.,][]{2000MNRAS.318L..65F} that do affect hardness ratio (see text and figures below). In the rest of the paper we ignore the regions that are known to be affected by the low energy absorption.

\subsection{Modified hardness ratio}
\label{sec:mhr}
An even simpler approach is to use hardness-ratio maps, based on the
images in two sufficiently different energy bands, which we designate below as $s$ and $h$, corresponding to energy intervals $[E_{s,1}:E_{s,2}]$  $[E_{h,1}:E_{h,2}]$.  Once again, the
assumption that the perturbations are small allows one to rewrite the
expression for the hardness ratio $H$ as
\begin{eqnarray}
  H(x,y)=\frac{I_h(x,y)}{I_s(x,y)}=\frac{I^0_h(x,y)\left (1+ \frac{\delta I_h(x,y)}{I^0_h(x,y)} \right )}{I^0_s(x,y)\left (1+\frac{\delta I_s(x,y)}{I^0_s(x,y)}\right )}\approx \nonumber \\
  \frac{I^0_h(x,y)}{I^0_s(x,y)}\left(1+\frac{\delta I_h(x,y)}{I^0_h(x,y)}- \frac{\delta I_s(x,y)}{I^0_s(x,y)}\right),
  \label{eq:hr}
\end{eqnarray}
where  $I_h$ and $I_s$ are the projected X-ray images in the
hard and soft band respectively:
\begin{eqnarray}
  I_{s,h}(x,y)=\int n^2 \Lambda_{s,h}(T) dz, \label{eq:i} \\
  \Lambda_{s,h}(T)=\int_{s,h} \epsilon(E,T) dE \label{eq:lam},
\end{eqnarray}
where $n=n(x,y,z)$ is the gas density; the integration in
eq.~(\ref{eq:i}) is over the line of sight (along the $z$ axis), while in
eq.~(\ref{eq:lam}) the integration is over the energy intervals,
corresponding to the $s$ or $h$ bands respectively. $I^0_h$ and $I^0_s$ are suitable simple
models, describing global radial profiles, and $\delta I_h=I_h-I^0_h$ and
$\delta I_s=I_s-I^0_s$ are the deviations of X-ray images from these simple
models. For instance, a $\beta$-model \citep{1978A&A....70..677C}
\begin{eqnarray}
I^0(x,y)=C\left [1+ \left ( \frac{\sqrt{x^2+y^2}}{r_c}\right )^2\right ]^{-3\beta+1/2}  
\end{eqnarray}
can be used as $I^0_h$ and $I^0_s$. For a cluster with a radial temperature gradient, not only the normalization $C$, but also other parameters, $\beta$ and $r_c$, of the $\beta$-models describing  $I^0_h$ and $I^0_s$ may be different.

One can see from eq.~(\ref{eq:hr}) that if $\displaystyle \frac{\delta
  I_h(x,y)}{I^0_h(x,y)} \ll 1$ and $\displaystyle \frac{\delta
  I_s(x,y)}{I^0_s(x,y)} \ll 1$, then the calculation of the hardness
ratio can be factorized into two terms $H=H_gH_p$ [see Fig.~\ref{fig:hrpt}), where the first term $\displaystyle H_g=\frac{I^0_h(x,y)}{I^0_s(x,y)}$ describes a global profile and
the second term $H_p$ [the term in parentheses in eq.~(\ref{eq:hr})] 
describes the variations of the hardness ratio associated with 
deviations of observed images from the global $I^0_h$ and $I^0_s$
models. This $H_p$ term can be evaluated as the difference between (adaptively) smoothed images $\displaystyle \frac{\delta I_h(x,y)}{I^0_h(x,y)}$ and $\displaystyle
\frac{\delta I_s(x,y)}{I^0_s(x,y)}$.
The resulting hardness ratio $H=H_gH_p$ is shown in the right
panel of Fig.~\ref{fig:tmap}.  Of course, if one is interested in the
temperature map, these values of $H$ have to be
converted\footnote{This conversion is straightforward, since the
  expected hardness ratio as a function of temperature can be easily
  predicted.} to $T$, unlike in the procedure described in the previous
section, where the temperature is obtained directly.

Notice that $H_p$ involves only a division of the data by a smooth
global model.  This means that a notorious problem of having noisy
data in the denominator is avoided by a suitable choice of global
models $I^0_h$ and $I^0_s$, subject to the condition that deviations
from these models are small. The implication is that we can avoid 
excessive smoothing of observed images needed to suppress noise in the expression $H=I_h/I_s$.

\section{ICM emissivity and projected X-ray images}
\label{sec:em}
In the above section we have factorized the hardness ratio $H=H_gH_p$
into ``global'' and ``perturbed'' parts. Below we use a similar
approach to single out and characterize the properties of the ICM
perturbations on small scales.

As a first step, we want to calculate the variations of the plasma volume emissivity in a given band $\displaystyle f=n^2\Lambda_B(T)$, due to the density and temperature variations. Here the index $B$ corresponds to one of the bands ($s$ or $h$).
For the purposes of this paper, we assume that the gas density $n$
and temperature $T$ distributions in a cluster can be factorized into
an ``unperturbed'' spherically symmetric model ($n_0(r)$ and $T_0(r)$) and
small-amplitude perturbations with respect to this model:
\begin{eqnarray}
n(x,y,z)=n_0(r)\times [1+\delta_n(x,y,z)] \label{eq:pert}\\
T(x,y,z)=T_0(r)\times [1+\delta_T(x,y,z)], \nonumber
\end{eqnarray}
where $\displaystyle r=\sqrt{x^2+y^2+z^2}$.  We further assume that
the relative perturbations $\delta_n$ and $\delta_T$ are not
independent, but associated with a particular type of
density/temperature perturbation characterized by the correlated changes of both quantities
\begin{eqnarray}
  \frac{d\ln T}{d\ln n}\equiv \alpha,
  \label{eq:alpha}
\end{eqnarray}
so the relation
between small perturbations $\delta_n$ and $\delta_T$ has the form $\displaystyle
\delta_T=\alpha\delta_n$.  Below we interpret perturbations with
different $\alpha$ as perturbations having different effective
equations of state (EoS). Note, that this is not necessarily the EoS in
the thermodynamic sense. By virtue of eq.~(\ref{eq:pert}) we probe
deviations of the gas properties from a smooth global model at a given
position in a cluster. Thus, the effective EoS characterizes the
fluctuations of gas properties for gas lumps located at the same
radial distance from the center (at least for spherically symmetric
models), rather than the changes of the thermodynamic properties of
the same gas lump.

In this language, $\alpha=\gamma-1$, where $\gamma$ is the effective adiabatic index of a perturbation when the pressure-density relation is $P\propto\rho^\gamma$. 
For the purpose of this paper, we restrict the set of perturbation types to adiabatic, isobaric and isothermal:
\begin{eqnarray}
\alpha&= 2/3 &~~ {\rm adiabatic,} ~~\gamma=5/3 \nonumber \\
\alpha&=-1 &~~ {\rm isobaric,} ~~\gamma=0 \\
\label{eq:proc}
\alpha&=0 &~~{\rm isothermal,} ~~\gamma=1. \nonumber
\end{eqnarray}
For example, adiabatic perturbations can be caused by sound waves
(weak shocks) going through the gas. Isobaric perturbations, associated
with entropy variations of the gas lumps in pressure equilibrium with
each other, could be due to slow displacement of fluid elements from
their equilibrium positions (e.g., gravity waves). Apparently ``isothermal''
perturbations could be due to bubbles of relativistic plasma, which
are devoid of X-ray emitting gas; if the pressure of the relativistic
plasma matches the ambient gas pressure, the bubbles will be seen as
X-ray-dim ``cavities'', suggesting a drop of thermal-gas density
without apparent changes in the gas temperature.

\begin{figure*}
  \begin{minipage}{0.49\textwidth}
    \includegraphics[trim= 0mm 0cm 0mm 0cm, width=1.\textwidth,clip=t,angle=0.,scale=1.0]{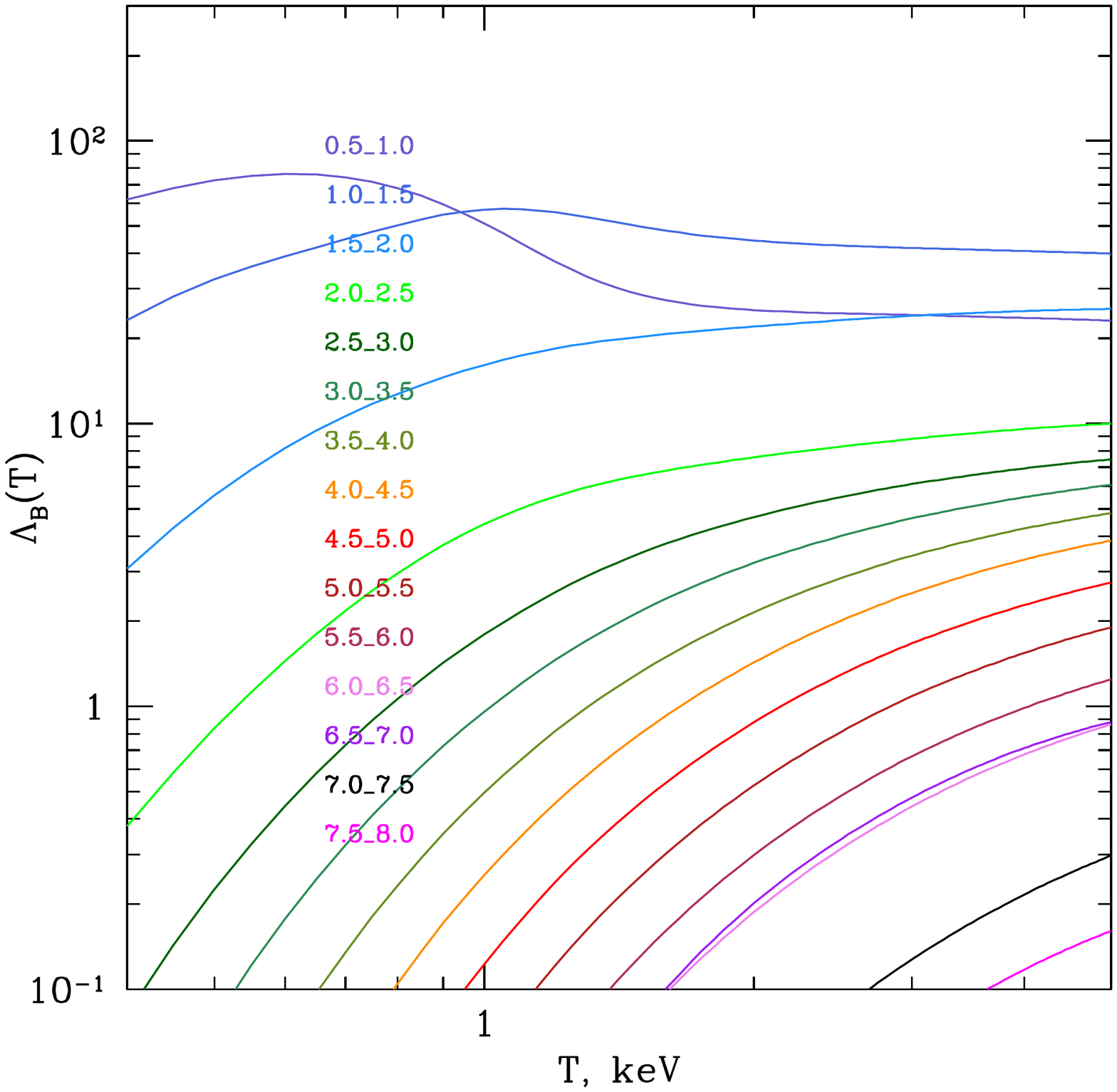}
  \end{minipage}
    \begin{minipage}{0.49\textwidth}
    \includegraphics[trim= 0mm 0cm 0mm 0cm, width=1.\textwidth,clip=t,angle=0.,scale=1.0]{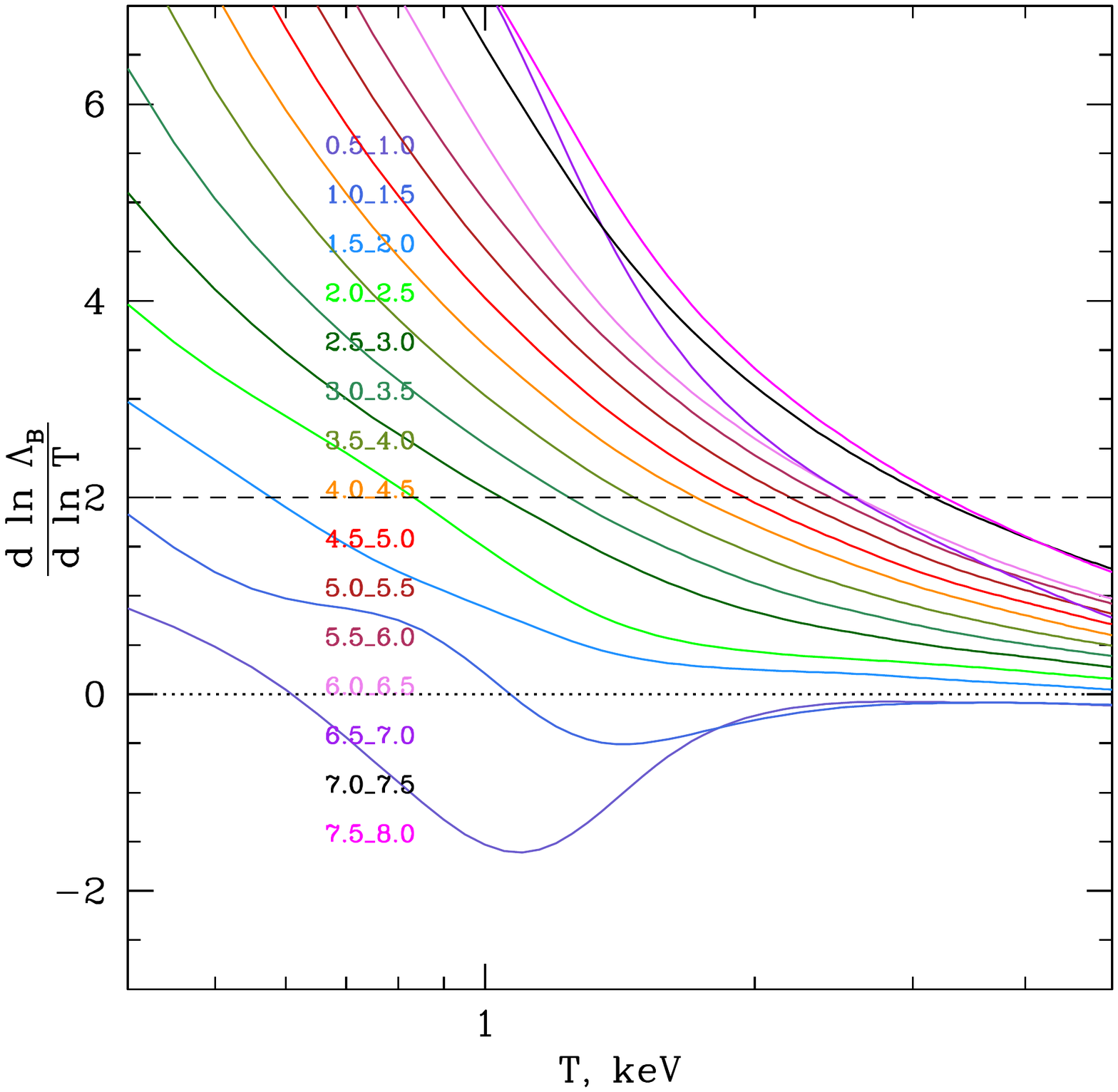}
    \end{minipage}
\caption{{\bf Left:} X-ray emissivity as a function of temperature. The
  emissivity (in arbitrary units) is calculated in half keV wide
  bands of ACIS-I (energy increases from top to bottom and is shown in
  the legend), using the APEC model of optically thin plasma with the
  abundance of heavy elements taken to be 0.4 solar.  {\bf Right:} Logarithmic
  derivative of the emissivity over temperature $\displaystyle
  \frac{d\ln \epsilon_B}{d\ln T}$ for the same set of bands as in the
  left panel. Energy increases from bottom to top. As expected, at
  temperatures $\ga 2$ keV the emissivity in the soft band (below
  $\la$2.5 keV) does not depend on temperature. For harder bands, this
  is not true. The dashed horizontal line shows for each band at what
  temperature $\displaystyle \frac{d\ln \epsilon_B}{d\ln T} \approx
  2$.  If this condition is satisfied, then pure isobaric
  perturbations will not be visible in this band \citep[][; see Figs, 10 and 11]{2007ApJ...665.1057F}. Note also that for low cluster temperatures ($T\lesssim1.5$~keV), the derivative is negative in the softest (0.5-1.0~ keV) band. This effectively  decreases the sensitivity of images in this band to adiabatic  to adiabatic perturbations (see eq.~[\ref{eq:ib})].
\label{fig:epsilon}
\label{fig:dlnedlnt}
}
\end{figure*}

As in \S\ref{sec:tmap} we use the APEC model
\citep{2001ApJ...556L..91S} as implemented in XSPEC
\citep{1996ASPC..101...17A} to derive the energy dependent plasma emissivity
$\epsilon(E,T)$, fixing the abundance of heavy elements to 0.4
solar. Unlike eq.~(\ref{eq:lam}), we now specialize the emissivity
$\Lambda_{B}(T)$ in a given energy band $B$ to the \Chandra ACIS-I
instrument. Namely, we convolve $\epsilon(E,T)$ with the ACIS-I
response (including effective area) and then integrate over energy
channels. Examples of the temperature-dependent emissivity $\Lambda_B(T)$ for several representative energy bands are shown in Fig.~\ref{fig:epsilon}. 

\begin{figure}
      \begin{minipage}{0.49\textwidth}
        \includegraphics[trim= 0mm 0cm 0mm 0cm, width=1.0\textwidth,clip=t,angle=0.,scale=1.0]{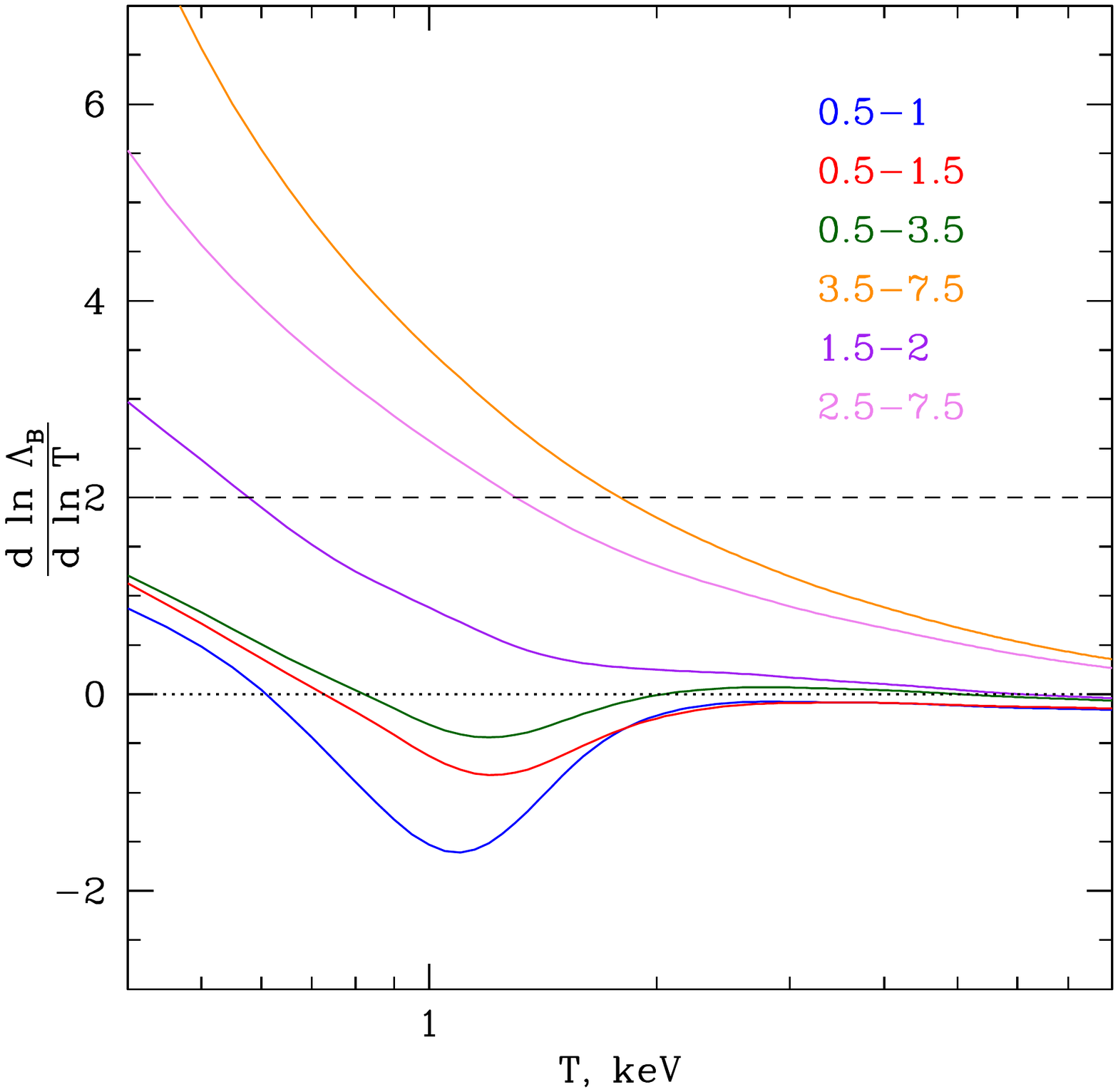}
      \end{minipage}
      \caption{The same as in the Fig.~\ref{fig:dlnedlnt}, but for broader energy bands.
        \label{fig:dlnedlnt_broad}
      }
\end{figure}

We can now predict the relative perturbation $\delta_f$ of the plasma volume emissivity (X-ray flux per cm$^{3}$) $f=n^2\Lambda_B(T)=f_0(1+\delta f)$ in a given \Chandra band $B$ for a particular type of
perturbation, characterized by the parameter $\alpha$ [see eq.~(\ref{eq:alpha})]
\begin{eqnarray}
  \delta_f&\approx&\left [2+  \frac{d\ln T}{d\ln n} \frac{d\ln \Lambda_B}{d\ln T}|_{T_0} \right ] \delta_n \nonumber \\
  &\approx&\left [2+ \alpha \frac{d\ln \Lambda_B}{d\ln T}|_{T_0} \right ] \delta_n.
  \label{eq:df}
\end{eqnarray}
Since the value $\displaystyle \frac{d\ln \Lambda_B}{d\ln T}$ is
known for each energy band (see Figs.~\ref{fig:dlnedlnt} and \ref{fig:dlnedlnt_broad}), for a given $\delta n$ the amplitude of the flux perturbations can be predicted for
each type of perturbation, i.e. for different $\alpha$. If the amplitude is
measured in two energy bands $B_1$ and $B_2$ (having different $\displaystyle \frac{d\ln \Lambda_B}{d\ln T}$) one can construct a linear
combination of the amplitudes that emphasizes (or suppresses)
a particular type of perturbation. The coefficients for such linear combinations can be derived from the expected ratio $R_f$ of fluxes in these two bands
\begin{eqnarray}
R_f= \frac{\delta_{f_2}}{\delta_{f_1}}=\frac{\left [2+ \alpha \left( \frac{d\ln
      \Lambda_{B_2}}{d\ln T}\right)|_{T_0} \right ]}{\left [2+ \alpha
    \left( \frac{d\ln \Lambda_{B_1}}{d\ln T}\right)|_{T_0} \right ]}.
  \label{eq:fratio}
\end{eqnarray}
A particular example of expected ratios for different types of
perturbations is shown in Fig.~\ref{fig:ratio} for $B_1=$0.5-3.5~keV
and $B_2=$3.5-7.5~keV.

Of course, observations provide us projected images in different
energy bands, rather than direct measurements of volume
emissivity/flux variations as in eq.~(\ref{eq:df}). Handling projected
images is discussed in the next section.

\subsection{Projected X-ray images}
An observed X-ray image in a given energy band $B$ is the projection of
the volume emissivity along the line of sight
\begin{eqnarray}
I_B(x,y)&=&\int f dz \\
\label{eq:ib}
&=&\int n_0^2 \Lambda_B(T_0) \left (1+\left [2+ \alpha \left( \frac{d\ln \Lambda_B}{d\ln T}\right) \right ] \delta_n\right ) dz. \nonumber
\end{eqnarray}
For an unperturbed gas distribution, i.e., when $\delta_n=0$, the above
expression provides a smooth model image in each band
\begin{eqnarray}
I_{B}^0(x,y)=\int n_0^2 \Lambda_B(T_0) dz. 
\end{eqnarray}

The ratio $J_B(x,y)$ of the observed image to the model image provides the
measure of surface brightness fluctuations relative to the model in a
given band
\begin{eqnarray}
J_B(x,y)=\frac{I_B(x,y)}{I_{B}^0(x,y)}-1.
\end{eqnarray}
Let us consider a small perturbation that is located at  $z=z_0$
along the line of sight, and has a small spatial extent $\Delta z$.  Then
\begin{eqnarray}
  J_B(x,y)=\frac{n_0^2(r) \Lambda_B\left (T_0\left (r\right )\right)\delta_n \Delta z \left [2+ \alpha \left( \frac{d\ln \Lambda_B}{d\ln T}\right)|_{T_0(r)} \right ]}{I_{B}^0(x,y)},
  \label{eq:j}
\end{eqnarray}
where $\displaystyle r=\sqrt{x^2+y^2+z_0^2}$. We now consider several
limiting cases.

\begin{figure}
  \begin{minipage}{0.49\textwidth}
    \includegraphics[trim= 0mm 0cm 0mm 0cm, width=1.0\textwidth,clip=t,angle=0.,scale=1.0]{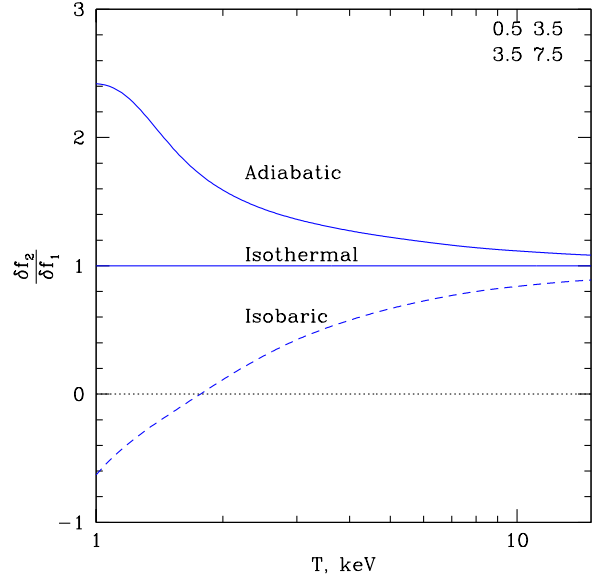}
  \end{minipage}
  \caption{Ratio of fluxes (volume emissivities) for different types of perturbations as
    a function of temperature for the 0.5-3.5 and 3.5-7.5 keV bands. The
    top curve is for adiabatic perturbations, predicting that the
    amplitude is larger in the hard band. The bottom curve is for
    isobaric perturbations, in which case the amplitude in the hard
    band is lower than in the soft band. In particular, the amplitude
    is close to zero in the 3.5-7.5 keV gas at $T\sim2$ keV. At lower
    temperatures, the hard-band amplitude even becomes negative,
    i.e., increasing density causes the 3.5-7.5 keV flux to drop. The
    solid horizontal line corresponds to isothermal perturbations, which produce the same 
    amplitude in both bands.
    \label{fig:ratio}
    }
\end{figure}

If the cluster is globally isothermal, then $T_0(r)={\rm const}$,
$\Lambda_B(r)={\rm const}$ and $\left [2+ \alpha \left( \frac{d\ln
    \Lambda_B}{d\ln T}\right)|_{T_0(r)} \right ]={\rm const}$. With
this assumption the ratio of amplitudes $J$ in two bands ($B_1$ and
$B_2$) at any point of the image is simply
\begin{eqnarray}
R(x,y)=R_f,
\label{eq:r}
\end{eqnarray}
where $R_f$ is give by eq.~(\ref{eq:fratio}), see also Fig.~\ref{fig:ratio}.
Notice that $R(x,y)$ does not depend on the position or the amplitude of the perturbation and is therefore a direct proxy for $\alpha$ -- the type of the perturbation. 

If the cluster is not isothermal, as is the case for typical cool-core
clusters, then the ratio $\displaystyle \frac{J_{B_2}}{J_{B_1}}$ will be position dependent. It is therefore useful to introduce an additional position-dependent
correction $\zeta_B(x,y)$ to eq.~(\ref{eq:j}) to at least partly mitigate the problem
\begin{eqnarray}
\tilde{J}_B(x,y)=J_B(x,y)\zeta_B(x,y),
\label{eq:rc}
\end{eqnarray}
so that in the ratio $\displaystyle \frac{\tilde{J}_{B_2}}{\tilde{J}_{B_1}}$ the position-dependence approximately cancels out.   

\begin{figure}
  \begin{minipage}{0.49\textwidth}
    \includegraphics[trim= 0mm 0cm 0mm 0cm, width=1.0\textwidth,clip=t,angle=0.,scale=1.0]{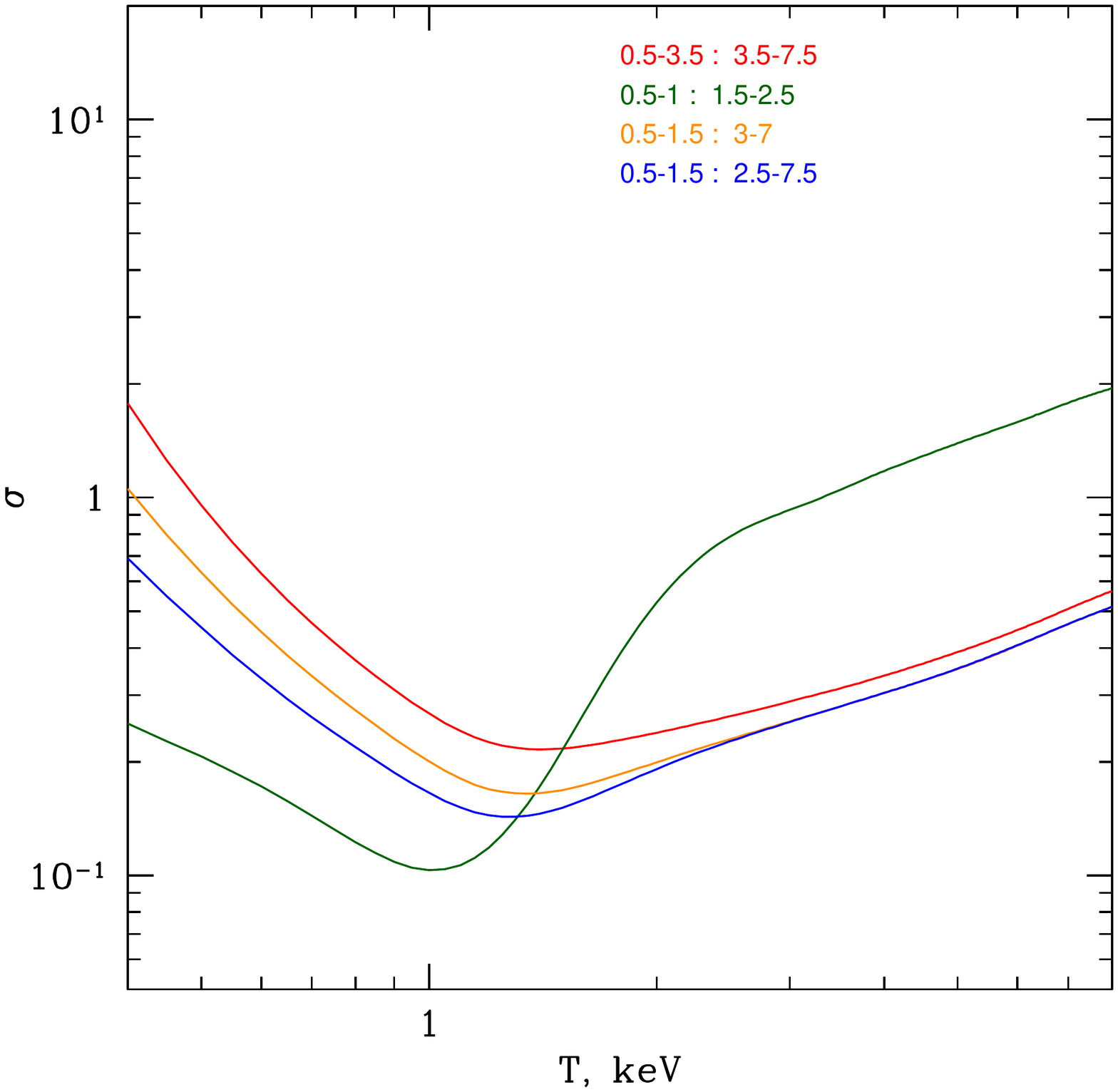}
  \end{minipage}
  \caption{Expected RMS ($\sigma$) of the manipulated image [see eq.~(\ref{eq:x})] with
    isothermal perturbations removed.     The uncertainty is calculated as
    a function of temperature for several pairs of energy bands (see
    legend). The optimal choice is achieved when $\sigma$ is lowest. It is clear that for $T\la 1$,
    keV the best choice is the pair of images in the 0.5-1.0 and 1.5-2.5 keV
    bands. At $T\ga 1$ keV, several other pairs (e.g., 0.5-1.5 and 2.5-7.5 keV or 0.5-3.5 and 3.5-7.5 keV)
    are expected to perform better. Note that only the statistical
    uncertainty arising from photon counting noise has been considered here.
    \label{fig:sigma}
    }
\end{figure}

The simplest correction comes from the assumption that the observed
perturbation is located close to the mid-plane of the cluster, i.e.,
at $z_0=0$. In this approximation,
\begin{eqnarray}
\zeta_B(x,y)\propto\frac{I_{B}^0(x,y)}{\Lambda_{B}(T_0(r))}.
\end{eqnarray}
From eq.~(\ref{eq:j}) it is clear that such corrections would recover the
desired property of the ratio $R(x,y)=\displaystyle
\frac{\tilde{J}_{B_2}}{\tilde{J}_{B_1}}$ to be sensitive only to the
value of $\alpha$.  Since the perturbations located close to the mid-plane are favoured by the $n^2$ dependence of the gas volume
emissivity, this correction factor is expected to work well. One can
then use $R(x,y)$ as a tool to classify individual perturbations
according to their effective EoS.

Alternatively, one can assume that perturbations are small and
volume-filling (i.e., that they are quasi-uniformly distributed over
the cluster volume) and calculate an averaged correction, weighted
with the $n_0^2$ term to account for the suppression of their contributions
to the image due to projection effects \citep[see,
  e.g.,][]{2012MNRAS.421.1123C}. Of course, in this case the
correction factor works only ``on average''. In practice, for
realistic temperature profiles, the two forms of the correction factor are
not very different, because $n_0^2$ term favours perturbtions located close to mid-plane.

\section{Arithmetic with X-ray images}
\label{sec:arith}

\subsection{Energy bands ``free'' from isobaric perturbations}
From eq.~(\ref{eq:df}), it is clear that, if, for a given $T$, the value of
$\displaystyle \frac{d\ln \Lambda_B}{d\ln T}$ is close
to $2$, then for isobaric perturbations (i.e., $\alpha=-1$) no flux
variations are expected. The dashed horizontal lines in the right panel of
Fig.~\ref{fig:dlnedlnt} and in Fig.~\ref{fig:dlnedlnt_broad} correspond to
$\displaystyle \left( \frac{d\ln \Lambda_B}{d\ln T}\right)=2$. For
instance, for the gas temperature $T\sim$2 keV the isobaric
perturbations should not be present in the images in the 3.5-7.5 keV
band (see Fig.~\ref{fig:dlnedlnt_broad}). This approach was used by
\citet{2007ApJ...665.1057F} for M87 ($T_0\sim 2$~keV) to avoid contamination of images by
prominent isobaric structures and to reveal quasi-spherical weak shocks
around the nucleus. In principle, for any given gas temperature one can try
to select an appropriate band, but the limited energy band accessible to Chandra makes
it difficult to handle hot clusters.

While it is possible to generate images free of isobaric
perturbations, for isothermal and adiabatic processeses this procedure
will not work\footnote{Note, however, that for $T\sim 1$ keV, the
contribution of adiabatic processes is expected to be severely attenuated in the
very soft energy band, where $\displaystyle  \frac{d\ln
  \Lambda_B}{d\ln T}$ is negative (see
Fig.~\ref{fig:dlnedlnt}).}. A more promising approach is to combine
two images in different energy bands, as is discussed below.

\subsection{Arithmetic with two images}
\label{sec:a2}
We now combine two flattened images $J_{B_1}$ and $J_{B_2}$ (or corrected images $\tilde{J}_{B_1}$ and $\tilde{J}_{B_2}$) in such a way that perturbations with a given effective EoS of state (i.e., given $\alpha$) are cancelled in the combined image
\begin{eqnarray}
  X=c_{B_1}J_{B_1}+c_{B_2}J_{B_2},
  \label{eq:x}
\end{eqnarray}
where coefficients $c_{B_1}$ and $c_{B_2}$ satisfy the condition
\begin{eqnarray}
  c_{B_1}\left [2+ \alpha_0 \left( \frac{d\ln
      \Lambda_{B_1}}{d\ln T}\right) \right ]+c_{B_2} \left [2+ \alpha_0 \left( \frac{d\ln
      \Lambda_{B_2}}{d\ln T}\right) \right ] =0,
  \label{eq:alpha0}
\end{eqnarray}
where $\alpha_0$ characterizes perturbations that we wish to remove.
It is useful to impose an additional constraint on $c_{B_1}$ and $c_{B_2}$
\begin{eqnarray}
  c_{B_1}\left [2+ \alpha_R \left( \frac{d\ln
      \Lambda_{B_1}}{d\ln T}\right) \right ]+c_{B_2} \left [2+ \alpha_R \left( \frac{d\ln
      \Lambda_{B_2}}{d\ln T}\right) \right ] =1,
  \label{eq:alphar}
\end{eqnarray}
where $\alpha_R\ne\alpha_0$ can be chosen arbitrarily.  Then the
amplitudes of perturbations having effective an EoS characterized by
$\alpha_R$ will reflect pure density variations associated with this particular type of
perturbation. As is wellknown (see \S\ref{sec:em} and Fig.~\ref{fig:epsilon}), in the soft band (e.g.,
0.5-2 keV) the observed perturbations are largely sensitive to
density perturbations, since the emissivity in this band weakly
depends on temperature. 
From eqs.~\ref{eq:j}, \ref{eq:x},
\ref{eq:alpha0}, \ref{eq:alphar}, specializing for simplicity to the
case of an isothermal cluster, i.e., $\displaystyle \frac{n_0^2(r)
  \Lambda_{B_1}\left (T_0\left (r\right )\right)}{I_{B_1}^0(x,y)}=
\frac{n_0^2(r) \Lambda_{B_2}\left (T_0\left (r\right
  )\right)}{I_{B_2}^0(x,y)}$, the resulting amplitude of
perturbations, characterized by a given $\alpha$ is
\begin{eqnarray}
  X=\frac{n_0^2(r) \Lambda_{B}\left (T_0\left (r\right )\right)}{I_{B}^0(x,y)} A_{\alpha}  \delta_n \Delta z,
  \label{eq:xr}
\end{eqnarray}
where
\begin{eqnarray}
  A_\alpha=\frac{\alpha-\alpha_0}{\alpha_R-\alpha_0}.
\end{eqnarray}
Comparison with eq.~(\ref{eq:j}) shows that in the manipulated image
$X$, the perturbations having $\alpha=\alpha_0$ are suppressed
($A_\alpha=0$), while the perturbations with $\alpha=\alpha_R$ have
$A_\alpha=1$, i.e., their amplitude reflects pure density
variations\footnote{Note, that in eq.~(\ref{eq:j}) the amplitude of flux caused by a pure density perturbation will be factor of 2 higher than in
  eq.~(\ref{eq:xr}). This is because we set the r.h.s. of
  eq.~(\ref{eq:alphar}) to 1, rather than 2.}. The values of the factor
$A_{\alpha}$ in eq.~(\ref{eq:xr}) for perturbations characterized by $\alpha$ (given
$\alpha_R$ and $\alpha_0$) are calculated in Table~\ref{tab:a}, using
our standard set of values $\alpha=\{2/3,0,-1\}$.

\begin{table}
  \caption{Amplitudes of the perturbations in the manipulated images for our choice of $\alpha_0$ (excluded process) and $\alpha_R$ (process with unit amplitude).
    \label{tab:a}
  }
  \begin{tabular}{r | r | r | r | r}
    \hline
    & & \multicolumn{3}{|c|}{$\alpha$} \\
    $\alpha_0$ & $\alpha_R$ & 2/3 & 0 & -1 \\
    \hline
    & & \multicolumn{3}{|c|}{Amplitude, $A_\alpha$} \\
        2/3 & -1  & 0 & 2/5 & 1 \\
    0   & -1  & -2/3& 0 & 1\\
    -1  & 2/3 & 1 & 3/5 & 0 \\
    \hline
  \end{tabular}
\end{table}

Note that the above procedure can eliminate from our images the
perturbations with any given EoS even if several perturbations with
different EoS's are projected on top of each other. On the other hand,
it does not provide an unambiguous identification of the EoS of the
remaining perturbations. One could extract the effective EoS directly from
eq.~(\ref{eq:r}), although in this case, the projection of several
overlapping perturbations makes the interpretation ambiguous.

\subsection{Selection of energy bands}
The selection of reference energy bands is driven by two competing
requirements: (i) to obtain the maximum number of counts in each energy band and
(ii) to separate the energy bands as much as possible to maximize the
difference in their response to different types of perturbations,
i.e., in the quantity $\displaystyle \frac{d\ln \Lambda_B}{d\ln T}$.  Assuming that
only pure photon counting noise is important, it is straightforward to
minimize the expected RMS of the final image $X$, given by eq.~(\ref{eq:x}), for
a particular choice of $\alpha_0$ and $\alpha_R$ \citep[see, e.g.,
  Appendix in][]{2016MNRAS.458.2902Z} by choosing appropriate energy bands. Since the amplitude of perturbations with $\alpha=\alpha_R$ does not depend on the choice of the energy band, such choice would maximize the signal-to-noise ratio for this type of perturbation. 
An example of optimal energy bands,
tuned for removal of isothermal perturbations, is shown in
Fig.~\ref{fig:sigma}. 

\subsection{Cross-power spectra}
\label{sec:cs}
Often a spherically symmetric $\beta$-model is clearly too simple to
describe the radial surface-brightness profile of a cluster, its ellipticity or any other
large-scale asymmetry. One possible way to remove these large-scale
asymmetries is to apply a high-pass filter to the images. In practice, the
simplest recipe is to start with a spherically symmetric $\beta$-model $I_{B}^0$,
divide the image $I_{B}$ by this model, smooth the resulting image and
multiply it back by the original $\beta$-model:
\begin{eqnarray}
I_{B}^1(x,y)=S\left [ \frac{I_{B}(x,y)}{I_{B}^0(x,y)}\right]I_{B}^0(x,y).
\end{eqnarray}
Here $S$ is a smoothing operator and $I_{B}^1(x,y)$ is the new global
model of the cluster that includes variations of
the cluster emission of the desired angular size (or larger).  Note
that this procedure preserves the global radial trend near the center as
long as the $I_{B}^0$ captures the trend. Of course one can use a more
complicated initial model for $I_{B}^0$, e.g., a double $\beta$-model,
or ellipsoidal models. 

The deviations of the surface brightness from the $I_{B}^1(x,y)$ now
contain only the perturbations on scales smaller that the width of
the smoothing filter. It is straightforward to extend this approach to
a fully scale-dependent analysis. Namely, one can calculate power
spectra, $P_1(k)$ and $P_2(k)$, and the cross-power-spectrum $P_{12}(k)$
of the images $J_1$ and $J_2$. Here $k$ is the wavenumber. Once $P_1$
and $P_2$ are corrected for the contribution of the photon counting
noise ($P_{12}$ is free from such noise, because the noise is independent in two images) one can construct two
scale-dependent quantities:
\begin{eqnarray}
  C(k)&=&\frac{P_{12}(k)}{\sqrt{P_{1}(k)P_2(k)}},\\
  R(k)&=&\frac{P_{12}(k)}{P_{1}(k)}.
\end{eqnarray}
The first quantity, $C(k)$,  is the coherence that shows how well
perturbations in one band correlate with perturbations in another
band. $C\approx 1$ means that the perturbations in one
band are linearly related to the perturbations in the other band (with
an arbitrary, but constant  proportionality coefficient across the image). In other words, perturbations with one particular effective EoS dominate. The second quantity, $R(k)$, is the
mean proportionally coefficient. If $C\approx 1$, then $R$
gives the coefficient for the dominant EoS [see eq.~(\ref{eq:r})].
If perturbations with
several different EoS's contribute significantly, then $R$ has an
intermediate value among the contributing EoS's.

We also emphasize that if we are interested only in the values of
$C(k)$ and $R(k)$, rather than the absolute normalization of the power
spectra, analysed images can be multiplied by an arbitrary large-scale
weighting function $w(x,y)$ (the same in both energy bands) before the
calculation of the power spectra. ``Large-scale'' in this context
means that we are interested in the values of $C(k)$ and $R(k)$ on
much smaller scales than those characteristic for $w(x,y)$.  The use
of such a weighting function might be useful to suppress excessive noise
in the intrinsically faint or underexposed regions of the images.

Thus, by calculating $C(k)$ and $R(k)$ one learns which
type (or types) of perturbations dominate at a given spatial
scale. This approach is useful for characterizing, in an objective way,
many weak structures, unlike the direct manipulation of images that
helps reveal only the most prominent structures directly visible in
the image. Such analysis was done for M87 \citep{2016ApJ...818...14A} and Perseus \citep{2016MNRAS.458.2902Z}.

\subsection{X-ray and SZ images}
The same approaches, outlined above, can be applied to a
combination of X-ray and Sunyaev-Zeldovich (SZ) data. An equivalent of
eq.~(\ref{eq:j}) for SZ is
\begin{eqnarray}
  J_{SZ}(x,y)=\frac{\sigma_T}{m_ec^2}\frac{n_0(r)kT_0(r)\delta_n \Delta z \left [1+ \alpha\right ]}{I_{SZ}^0(x,y)},
  \label{eq:jsz}
\end{eqnarray}
where $I_{SZ}^0(x,y)$ is the model of the global map of the Comptonization parameter; $\sigma_T$ is the Thomson scattering cross section; $k$, $m_e$ and $c$ are the Boltzmann constant, the electron mass and the speed of light, respectively. One can use eq.~(\ref{eq:jsz}) to get an expression for $\displaystyle  d_{SZ}=\delta_n \Delta z \left [1+ \alpha\right]$ and do the same for $\displaystyle d_{X}= \delta_n \Delta z \left [2+ \alpha \left( \frac{d\ln \Lambda_B}{d\ln T}\right)|_{T_0(r)} \right ]$ using eq.~(\ref{eq:j}). The relation between $d_{SZ}$ and $d_X$ is
\begin{eqnarray}
  d_{X}&=&K d_{sz},~{\rm where} \nonumber \\
  K&=&\frac{n_0(r)\Lambda_B\left [T_0(r)\right ]}{\frac{\sigma_T}{m_ec^2}kT_0(r)}\frac{I_{SZ}^0(x,y)}{I_{B}^0(x,y)}.
\label{eq:K}  
\end{eqnarray}
Notice, that even for an isothermal cluster the dependence on $r$ does
not cancel out because of the different dependence on $n_0$ of the
X-ray and SZ signals. If, for a prominent feature in the image, the
value of $\alpha$ is already known from X-ray analysis, then this
relation can be used to determine the location $z$, of the feature along the line of sight, since $z$ enters eq.~(\ref{eq:K}) via $r=(x^2+y^2+z^2)^{1/2}$ both in the
r.h.s. and l.h.s. of the equation.  Alternatively, if one can make a
guess on $z$, one can generate an X-ray image free from isobaric
perturbations (see \S\ref{sec:a2}) and then directly compare SZ and
X-ray images. This approach can be used to, e.g., differentiate
between the thermal or non-thermal nature of the gas providing pressure
support for AGN-inflated bubbles, or to prove that
pressure perturbations are due to sound waves. In the latter case,
an additional manipulated X-ray image, free from sound waves, will be
useful.

\section{Illustrative examples}
\label{sec:per}
To illustrate the approaches outlined above, we use Chandra observations of the Perseus and M87/Virgo clusters. The advantage of using these data sets is two-fold: (i) these are
the X-ray-brightest clusters in the sky and (ii) many features have already been
 identified by detailed analysis \citep[e.g.,][]{1993MNRAS.264L..25B,2000A&A...356..788C,2003MNRAS.344L..43F,2007ApJ...665.1057F,2016ApJ...818...14A,2016MNRAS.458.2902Z}.

\begin{figure*}
  \begin{minipage}{0.99\textwidth}
    \includegraphics[trim= 0mm 0cm 0mm 0cm, width=1\textwidth,clip=t,angle=0.,scale=
  1.0]{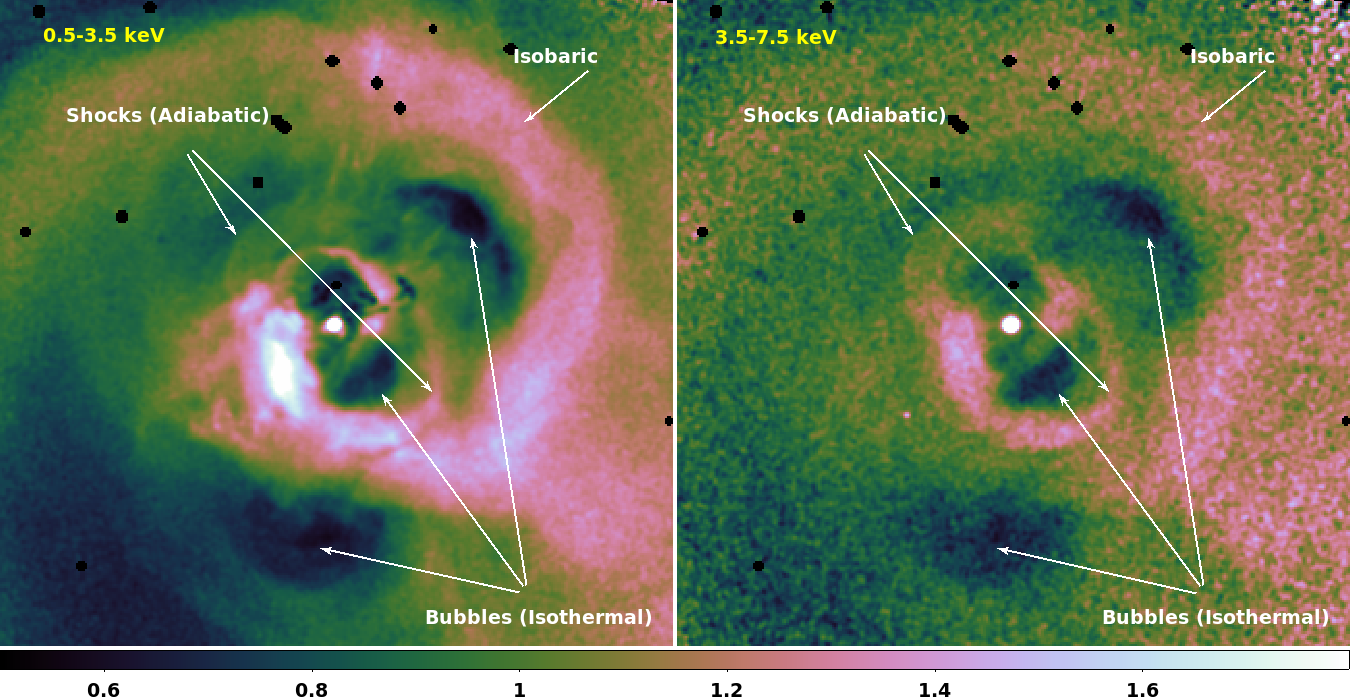}
  \end{minipage}
\caption{X-ray images ($5'\times 5'$; $\sim 100\times 100$~kpc) of the Perseus cluster core divided by their
  respective best-fitting beta-models. Left: 0.5-3.5 keV, right:
  3.5-7.5 keV. Labels mark several prominent features, which have been
  tentatively identified as shocks, bubbles and isobaric structures \citep[see, e.g.,][]{2011MNRAS.418.2154F}.
\label{fig:dbs}
}
\end{figure*}

The Chandra images of the Perseus cluster in the 0.5-3.5 and 3.5-7.5
keV bands, divided by their respective best-fitting $\beta$-models are
shown in Fig.~\ref{fig:dbs}. In the notation of \S\ref{sec:arith},
these are $\tilde{J}_B$ images [see eq.~(\ref{eq:rc})]. Labels in
Fig.~\ref{fig:dbs} mark several prominent features, which have been
tentatively identified as shocks, bubbles and isobaric structures
\citep[e.g.,][]{2003MNRAS.344L..43F,2016MNRAS.458.2902Z}. Those
identifications are based either on the comparison of X-ray and radio
images or on a detailed spectral analysis.

\begin{figure*}
  \begin{minipage}{0.99\textwidth}
    \includegraphics[trim= 0mm 0cm 0mm 0cm, width=1\textwidth,clip=t,angle=0.,scale=
  1.0]{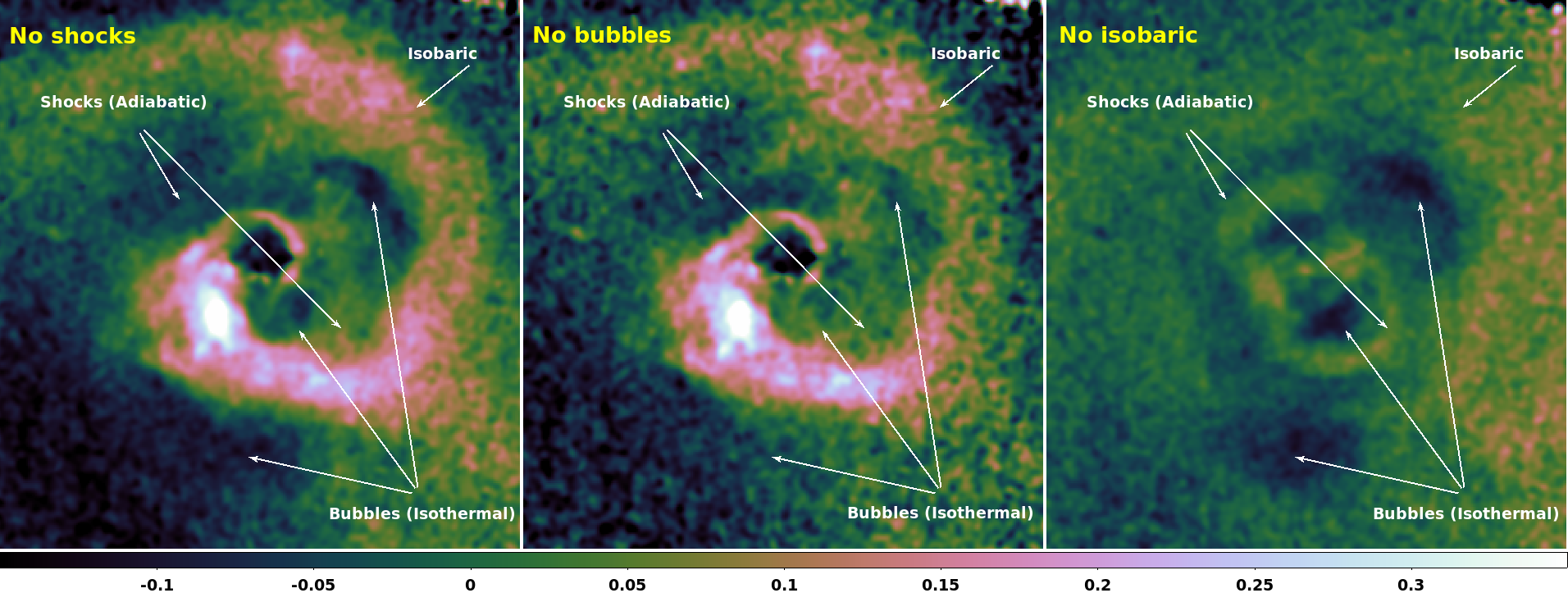}
  \end{minipage}
\caption{Manipulated X-ray images, based on the two images shown in
  Fig.~\ref{fig:dbs}. These two images were scaled and combined so as to exclude a
  particular type of perturbation. Other types of perturbations with
  different effective equations of state parametrized by
  $\alpha=\frac{d\ln T}{d\ln n}$, will still be
  present. In terms of our standard set of possible perturbations
  $\alpha=\{3/2,0,-1\}$, the expected amplitudes $A$ are: left panel,
  ``no shocks'', $A=\{0.0,0.4,1.0\}$; middle panel, ``no bubbles'',
  $A=\{-0.7,0.0,1.0\}$; right image, ``no isobaric structures'',
  $A=\{1.0,0.6,0.0\}$. One can see that some of the features prominent
   in Fig.~\ref{fig:dbs} disappear from one of the
  panels, suggesting that the density and temperature fluctuations obey
  a particular effective EoS. A feature to the West from the nucleus in the right panel is caused by cold gas absorption \citep[see, e.g.,][]{2000MNRAS.318L..65F}.
\label{fig:xia}
}
\end{figure*}

These images have been combined, using eq.~(\ref{eq:x}), into a new
set of images (Fig.~\ref{fig:xia}) to remove adiabatic, isothermal and
isobaric structures. As expected, proper scaling of relative weights
in a linear combination suppresses the substructures characterized by
a target EoS. The most striking is the case when isobaric structures
are removed (Fig.~\ref{fig:xia} right panel). Those structures
strongly dominate the original images (Fig.~\ref{fig:dbs}), especially
in the softer band, but are completely gone in the manipulated
image. The resulting image clearly shows a very symmetric
figure-8-like feature, which is believed to be due to compressed gas
around growing radio lobes, produced by the central AGN. The
figure-8-like shape is due to projection of two nearly circular
structures on either side of the nucleus. The ``older'' bubbles are
also clearly seen to the North-West and to the South of the
center. While most the above features have been seen before,
Fig.~\ref{fig:xia}, in addition, suggests that there is an envelope of
decreased thermal pressure in between the inner lobes (the
figure-8-like structure) and older bubbles. It likely that this
envelope is due to relativistic plasma occupying a fraction of volume
around the inner lobes. 

\begin{figure}
  \begin{minipage}{0.49\textwidth}
    \includegraphics[trim= 0mm 0cm 0mm 0cm, width=1.0\textwidth,clip=t,angle=0.,scale=1.0]{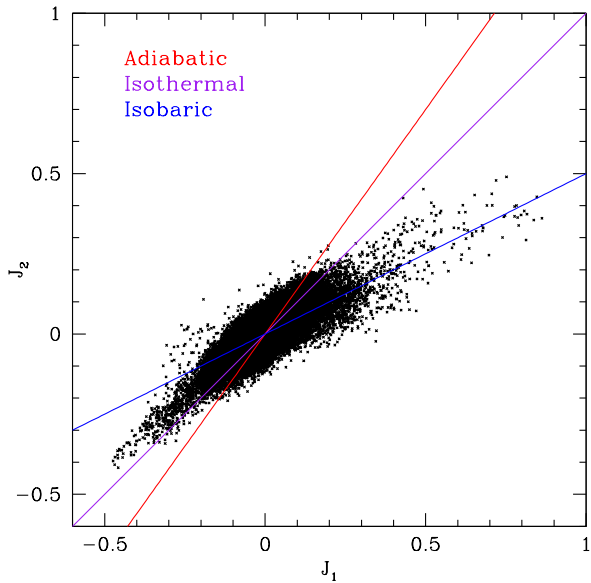}
  \end{minipage}
  \caption{Correlation between amplitudes of perturbations in the hard
    and soft bands for X-ray images of the Perseus cluster. Three lines show expected dependence for
    adiabatic, isothermal and isobaric perturbations. Clearly, large
    positive deviations correspond to isobaric perturbations (cool and
    dense structures), while large negative deviations are almost
    isothermal (bubbles). For this figure a  $16'\times16'$ image was used,
    i.e. larger than the images shown in Fig.~\ref{fig:dbs}.
    \label{fig:jcor}
  }
\end{figure}

Another way to characterize the observed fluctuations is to 
correlate the amplitudes in $J_1(x,y)$ and $J_2(x,y)$ pixel by pixel directly (see
Fig.~\ref{fig:jcor}). The three lines shown in Fig.~\ref{fig:jcor}
correspond to the expected correlation between fluxes in the soft and hard bands for adiabatic, isothermal
and isobaric fluctuations, respectively. As is clear from this figure, all large
positive perturbations correspond to isobaric perturbations (blue
line). In the original images (Fig.~\ref{fig:dbs}), those perturbations
correspond to the prominent spiral-like structure. At the same time,
large negative deviations follow an isothermal EoS. These perturbations
correspond to X-ray cavities (radio bubbles), visible in
Fig.~\ref{fig:dbs} as dark patches.

To study smaller-amplitude perturbations that are too weak to be identified
individually (given the noise in the image), a cross-spectrum approach
(\S\ref{sec:cs}) is a better option to characterize the mean
correlation between the perturbations. Moreover,  scale-dependent
nature of the cross-spectrum analysis helps avoid the impact of larger-scale asymmetries on the
resulting correlations. This approach was followed for the Perseus
cluster by \citet{2016MNRAS.458.2902Z} and for M87 by
\citet{2016ApJ...818...14A} and confirmed that isobaric perturbations
dominate the overall energy budget associated with perturbations.
Since all three manipulated images in Fig.~\ref{fig:xia} are shown in
the same color scale, one can immediately see that in terms of
variance, isobaric fluctuations clearly dominate. However, in order to
estimate the energy associated with the perturbations one has either
to assume a particular geometry, as was done for bubbles in many
studies, or to assume that the power spectrum of the
perturbations in 3D is isotropic. In the latter case, the power
spectrum analysis recovers the correct normalization of the 3D power
spectrum and can be used to estimate the total energy in the
perturbations \citep[see][for
  details]{2016ApJ...818...14A,2016MNRAS.458.2902Z}.

\begin{figure*}
  \begin{minipage}{0.99\textwidth}
    \includegraphics[trim= 0mm 0cm 0mm 0cm, width=1\textwidth,clip=t,angle=0.,scale=
  1.0]{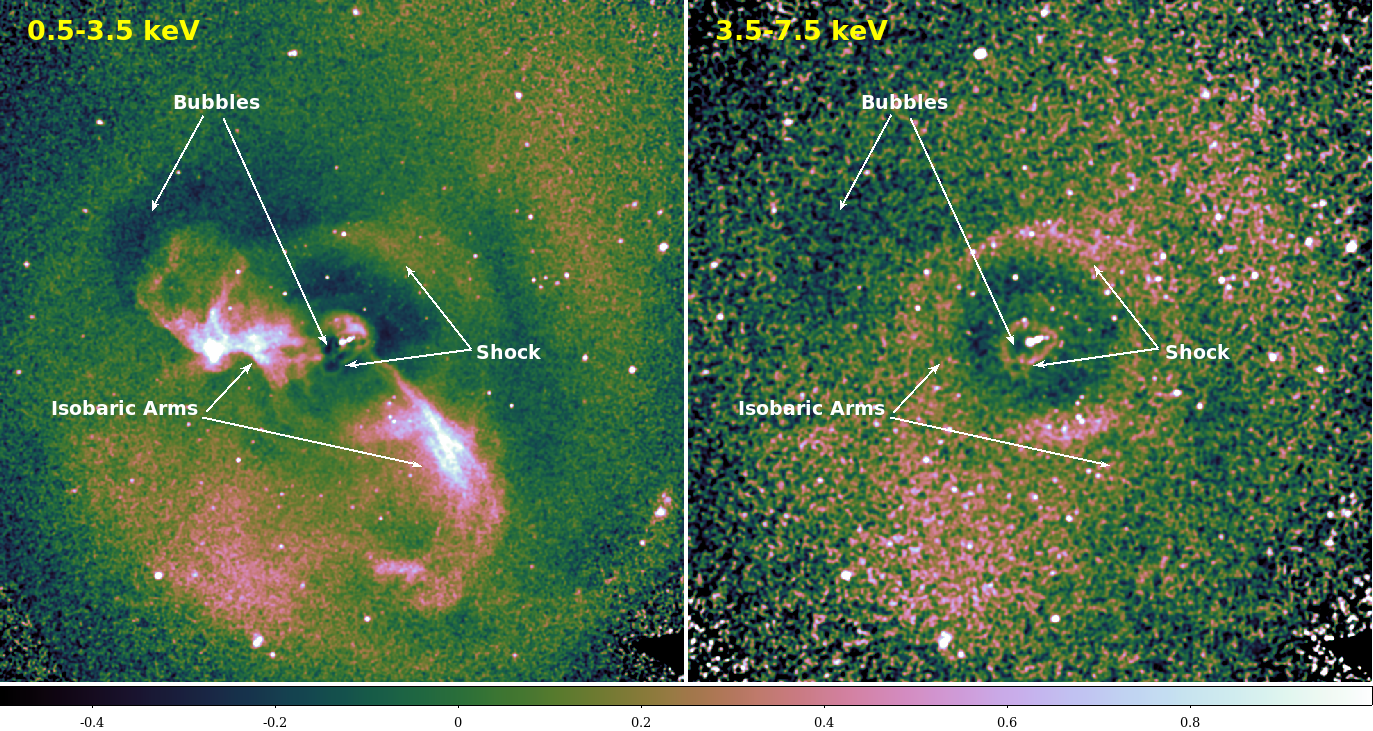}
  \end{minipage}
\caption{X-ray images ($16'\times 16'$; $\sim 75\times 75$~kpc) of the M87 core divided by their
  respective best-fitting $\beta$-models \citep[see, e.g.,][]{2005ApJ...635..894F,2007ApJ...665.1057F}. Left: 0.5-3.5 keV, right:
  3.5-7.5 keV. Similar images have already been shown in \citet{2016ApJ...818...14A}. Here we present them again to facilitate comparison with manipulated imaged of M87, shown in Fig.~\ref{fig:xia_m87}.
  \label{fig:dbs_m87}
  }
\end{figure*}

\begin{figure*}
  \begin{minipage}{0.99\textwidth}
    \includegraphics[trim= 0mm 0cm 0mm 0cm, width=1\textwidth,clip=t,angle=0.,scale=
  1.0]{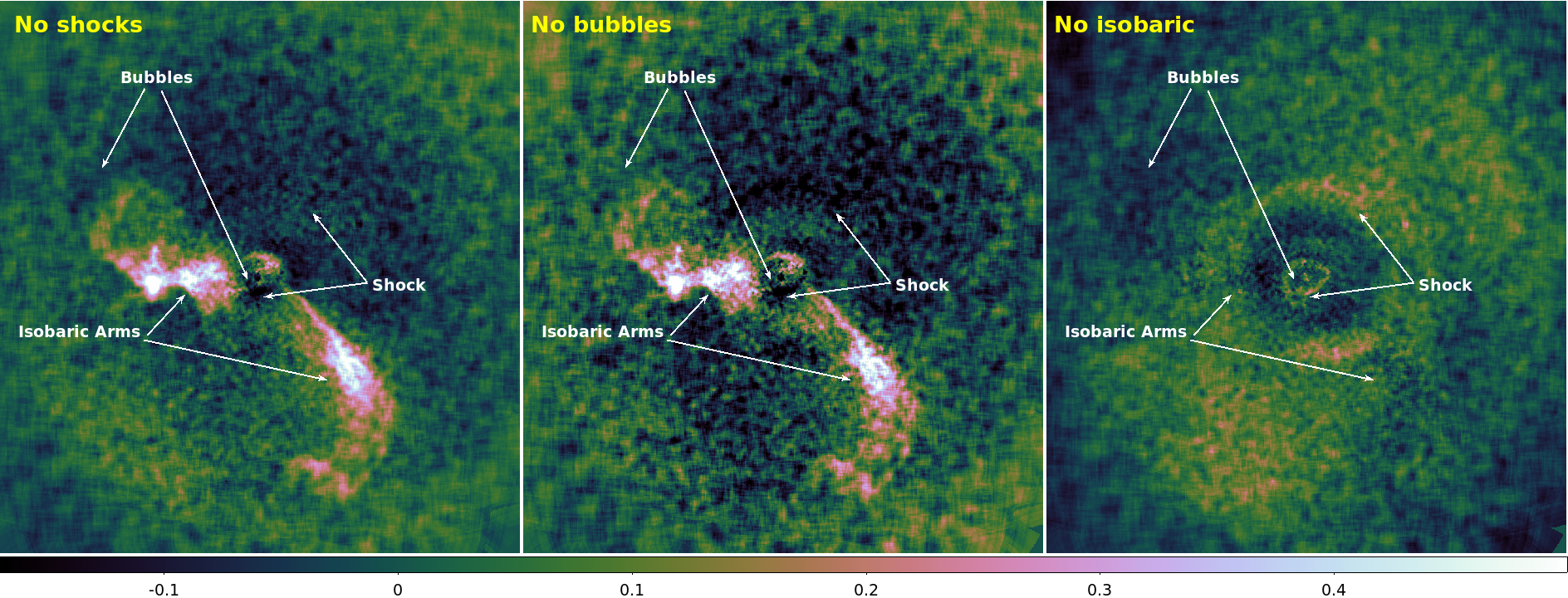}
  \end{minipage}
\caption{Manipulated X-ray images, based on the two images shown in
  Fig.~\ref{fig:dbs_m87}. The procedure used to generate the
  manipulated images is the same as in Fig.~\ref{fig:xia}. Since the
  characteristic temperature in M87 ($T\sim 2$~keV) is lower than in
  Perseus, the 3.5-7.5 keV image by itself is expected to be free from
  isobaric perturbations (see Fig.~\ref{fig:ratio}). As the result the
  manipulated image in the right panel looks similar to the original
  3.5-7.5 keV image in Fig.~\ref{fig:dbs_m87}.
\label{fig:xia_m87}
}
\end{figure*}
Fig.~\ref{fig:dbs_m87} and \ref{fig:xia_m87} show a similar analysis for
M87 images. As discussed by \citet{2007ApJ...665.1057F} the 3.5-7.5
keV band image of M87 (Fig.\ref{fig:dbs_m87}, right) by itself
provides a projected thermal pressure map (more accurately, projected
square of the pressure). From Fig.~\ref{fig:ratio} it is clear that for
the characteristic gas temperature in M87, $T\sim 2$~keV, the response to
isobaric perturbations in this energy band is indeed close to zero.
It is therefore not surprising that the manipulated image free from
isobaric perturbations (Fig.\ref{fig:xia_m87}, right) look very
similar to the original 3.5-7.5 keV band image.

Inspection of Figs.~\ref{fig:xia} and \ref{fig:xia_m87} suggests that,
in terms of the density perturbation amplitude, the isobaric structures
dominate, followed by bubbles (may still be isobaric, but with a
significant contribution to pressure from either relativistic
particles or very hot gas) and weak shocks. Such a hierarchy is best explained
in a ``slow'' AGN feedback scenario, when much of the mechanical
energy output of a central black hole is captured by the bubble
enthalpy that is gradually released during buoyant rise of the bubble
\citep[e.g.,][]{2000A&A...356..788C,2000ApJ...534L.135M}.

\section{Discussion and conclusions}
\label{sec:dis}

We have presented a set of  methods that help reveal the
(thermodynamic) nature of the gas perturbations, observed in relaxed
galaxy clusters. A modified hardness-ratio approach separates global
variations of the projected temperature and small-scale
substructure. As a result one can avoid excessive noise in the
hardness-ratio maps, since a division of observed images, which are
often noisy, is replaced by a subtraction of properly scaled images. An
extension of the same technique to the ``image arithmetic'' works best
for prominent structure and for datasets with excellent statistics,
visualizing the perturbations with a given effective equation of
state. For a global statistical characterization of many small
perturbations, the cross-power-spectrum approach is more appropriate
\citep[see][for details]{2016ApJ...818...14A,2016MNRAS.458.2902Z}. All
the methods that have been proposed are easy to implement and
computationally fast.

The above analysis makes two main simplifying assumptions. Firstly, we
assume that one can make a good guess about the ``unperturbed''
analytic model, since perturbations are calculated relative to this
model.  If the perturbations are on small scales, then an equivalent
assumption is that the unperturbed model is very smooth on the same
scales. Secondly, it is assumed that all perturbations have small
amplitudes, so all terms of order $\displaystyle \delta_n^2$ can be
neglected. While there are always modest departures from these
assumptions, the comparison of the manipulated images with the radio
data and with the results of detailed spectral analysis suggests that
this approach successfully classifies the types of perturbations and
helps to reveal their nature.

\FloatBarrier

\section{Acknowledgements}
EC acknowledges support by grant No. 14-22-00271 from the Russian Scientific Foundation. WF and CJ acknowledge support from NASA contract NAS8-03060 and the Smithsonian Institution. PA acknowledges support from Fondecyt grant 1140304

\label{lastpage}
\end{document}